\documentclass[aps,prl,superscriptaddress]{revtex4}
\usepackage{bm}
\usepackage{graphicx,amsmath,latexsym,amssymb}
\usepackage{color}
\usepackage{dcolumn}

\begin{document}

\definecolor{olive}{rgb}{0,1,.4}

\title{Observation of the intrinsic Abraham force in
time-varying magnetic and electric fields}
\author{G.L.J.A. Rikken }
\affiliation{Laboratoire National des Champs Magn\'{e}tiques Intenses\\
UPR3228 CNRS/INSA/UJF/UPS, Toulouse \& Grenoble, France.}
\author{B.A. van Tiggelen}
\affiliation{Univ. Grenoble 1 / CNRS, LPMMC UMR 5493, BP 166, 38042 Grenoble, France}

\date{\today}

\begin{abstract}

The Abraham force exerted by a time-dependent electromagnetic field on
neutral, polarizable matter has two contributions. The one induced by a
time-varying magnetic field and a static electric field is reported here for
the first time. We discuss our results in the context of the radiative
momentum in matter. Our observations are consistent with Abraham's and
Nelson's versions for radiative momentum.
\end{abstract}

\maketitle

\volumeyear{year}
\volumenumber{number}
\issuenumber{number}
\eid{identifier}
\date{\today }
\startpage{1}
\maketitle

The Abraham force is the force exerted by a time-dependent electromagnetic
field on neutral, polarizable matter, and has been debated for over a
hundred years. The macroscopic Maxwell's equations provide a continuity
equation for electromagnetic momentum that takes the general form%
\begin{equation}
\partial _{t}\mathbf{G}+\nabla \cdot \mathbf{T}=\mathbf{f}_{1}  \label{G}
\end{equation}%
with $\mathbf{G}$ \textquotedblleft some" electromagnetic momentum density, $%
\mathbf{f}_{1}$ \textquotedblleft some" force density exerted on the
radition, and $\mathbf{T}$ \textquotedblleft some" stress tensor. The
apparent arbitrariness in assigning expressions for $\mathbf{G}$, $\mathbf{T}
$ and $\mathbf{f}_{1}$ is known as the Abraham-Minkowski (AM)
controversy.(for reviews see e.g. \cite{Brevik1} and \cite{Pfeiffer}). In
the Minkowski version one adopts $\mathbf{G}=\mathbf{G}_{M}=\mathbf{D}\times
\mathbf{B}$ and Maxwell's equations lead to $\mathbf{f}_{1}(M)=\varepsilon
_{0}\left( \nabla \varepsilon _{r}\right) \mathbf{E}^{2}+\mu _{0}^{-1}\left(
\nabla \mu _{r}^{-1}\right) \mathbf{B}^{2}$, i.e. there is no force
proportional to both electric and magnetic field. The Abraham version
insists on a momentum proportional to the energy flow so that $\mathbf{G}=%
\mathbf{G}_{A}=\varepsilon _{0}\mu _{0}\mathbf{E}\times \mathbf{H}$ in which
case the Abraham force density takes the form $\mathbf{f}_{1}(A)=\mathbf{f}%
_{1}(M)-\varepsilon _{0}(\varepsilon _{r}-1/\mu _{r})\partial _{t}(\mathbf{E}%
\times \mathbf{B})$. The second term on the right hand side will be referred
to as the \textquotedblleft Abraham force density". Other versions can be
found in literature such as Peierls' proposition $\mathbf{f}%
_{1}(P)=\varepsilon _{0}(\varepsilon _{r}-1)({\partial _{t}\mathbf{E}}\times
\mathbf{B}-\frac{1}{5}\mathbf{E}\times \partial _{t}{\mathbf{B}})$ in which
case the full time derivative of the Abraham version has disappeared \cite%
{Peierls}. In the Einstein-Laub version the force $\mathbf{f}_{1}$ achieves
a gradient, i.e. a part of $\nabla \cdot \mathbf{T}$. For the homogenous
case, which we will consider for the remainder, the Einstein-Laub version is
equivalent to the Abraham version.

In this work we will measure explicitly the force exerted on matter by a
combination of a time-dependent electric field $\mathbf{E(}t\mathbf{)}$ and
a time-dependent magnetic field $\mathbf{B(}t\mathbf{)}$. The Newton-Lorentz
force on an object with mass density $\rho $ is unambiguously given by $\rho
\partial _{t}\mathbf{v}=\rho _{q}\mathbf{E}+\mathbf{J}_{q}\times \mathbf{B}$%
, with $\rho _{q}$ the charge density, $\mathbf{J}_{q}$ the charge current
density and $\mathbf{v}$ the velocity. Under the assumption of macroscopic
fields and in the absence of free charges and currents \cite{Jackson} it can
be cast in the form \cite{CommentPRL}%
\begin{equation}
\rho {\partial _{t}{\mathbf{v}}}+\nabla \cdot \mathbf{W}=\mathbf{f}_{2}.
\label{NL}
\end{equation}%
with $\mathbf{W}$ another stress tensor and $\mathbf{f}_{2}=-\mathbf{f}%
_{1}(M)+\varepsilon _{0}(\varepsilon _{r}-1)\partial _{t}(\mathbf{E}\times
\mathbf{B})$. Since $\mathbf{W}$ vanishes outside the object one identifies $%
\mathbf{F}=\int \mathbf{f}_{2}dV$ as the force exerted by the
electromagnetic field on the matter. From Newton's third law one expects
that $\mathbf{f}_{1}=-\mathbf{f}_{2}$. This is not valid for neither the
Minkowski nor the Peierls version, and true only for the Abraham version if $%
\mu _{r}=1$. The version for which Newton's third law strictly holds is the
Nelson version \cite{nelson}, in which the radiative momentum density is
chosen as $\mathbf{G}=\mathbf{G}_{N}=\varepsilon _{0}\mathbf{E}\times
\mathbf{B}$, i.e. equal to the expression in vacuum.

When taking a quantum-mechanical approach to the problem, new controversies
seem to appear. It is well-known that the presence of magnetic fields makes
the kinetic momentum $\mathbf{P}_{\mathrm{kin}}=\sum_{i}m_{i}\mathbf{\dot{r}}%
_{i}$ different from the conjugated momentum $\mathbf{P}=\mathbf{P}_{\mathrm{%
kin}}+\frac{1}{2}\sum_{i}q_{i}\mathbf{B}\times \mathbf{r}_{i}$. This
difference was recently put forward as a solution to the AM controversy \cite%
{Barnett}. However, in quantum mechanics a third \textquotedblleft
pseudo-momentum" $\mathbf{K}=\mathbf{P}_{\mathrm{kin}}+\sum_{i}q_{i}\mathbf{B%
}\times \mathbf{r}_{i}$ appears that is conserved in time \cite{Kawka}, and
clearly different from the other two. It is easy to verify that the validity
of the Nelson version is directly related to the conservation of $\mathbf{K}$%
. A second controversy was initiated by Feigel. \cite{Feigel}, arguing that
the quantum vacuum provides an additional contribution to the Abraham force.
This proposition was refuted theoretically \cite{Kawka} and experimentally
\cite{RikkenVanTiggelen}.

\begin{figure}[t]
\begin{center}
\resizebox{0.25\textwidth}{!}{%
\includegraphics{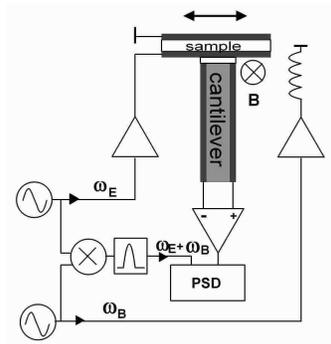}
}
\end{center}
\caption{Schematic setup of the experiment. A magnetic field B is applied
perpendicular to the drawing at frequence $\omega _{B}$, and high voltage at
frequency $\omega _{E}$\ is applied to the electrodes of the sample. The
cantilever signal is phase sensitively detected at the sum frequency $\omega
_{B}+\omega _{E}$. Typical sample size is 9 x 2 x 0,35 mm$^{3}.$}
\end{figure}

The experimental observation of the Abraham force induced by an \emph{%
oscillating} electric field and a \emph{static} magnetic field was reported
by James \cite{James} and by Walker et al \cite{Walker,Walker2} in solid
dielectrics, and recently by Rikken and Van Tiggelen in gazes \cite%
{RikkenVanTiggelen}. These observations clearly invalidated the Minkowski
version for $\mathbf{f}_{1}$ although modifications of the Minkowsi
energy-momentum tensor have been proposed to make it consistent with these
results \cite{Israel,Obukov}. However, the Abraham force due to an \emph{%
oscillating} magnetic field and a \emph{static} electric field has so far
never been observed and was even reported \emph{unobservable} \ in a
specifically designed experiment \cite{Walker3,Walker4}. The two cases
clearly correspond to physically different situations. A time-dependent
electric field creates moving charges ($\mathbf{J}_{q}=\partial _{t}\mathbf{P%
})$ that are subject to the Lorentz force. A time-varying magnetic field
induces a rotational electric field that acts on the polarization charges.
Walker et al suggested an explanation for their failure to observe the $%
\mathbf{E\times }\partial _{t}\mathbf{B}$\textbf{\ }component \cite{Walker5}%
, involving the compensation of this bulk component by a surface
contribution. The current experimental situation around the AM controversy
for low-frequency electromagnetic fields is therefore unsatisfactory as not
a single prediction for the Abraham force is experimentally confirmed. At
optical frequencies the situation is even less clear. Here, the Abraham
force cannot be observed directly, as it averages to zero over one cycle,
and one has to resort to the observation of momentum transfer from light to
matter, with the aforementioned conceptual difficulty which type of momentum
to consider. For recent discussions, see the two comments \cite{commentsshe}
on the work of She et al \cite{she}.

\begin{figure}
\begin{center}
\resizebox{0.5\textwidth}{!}{%
\includegraphics{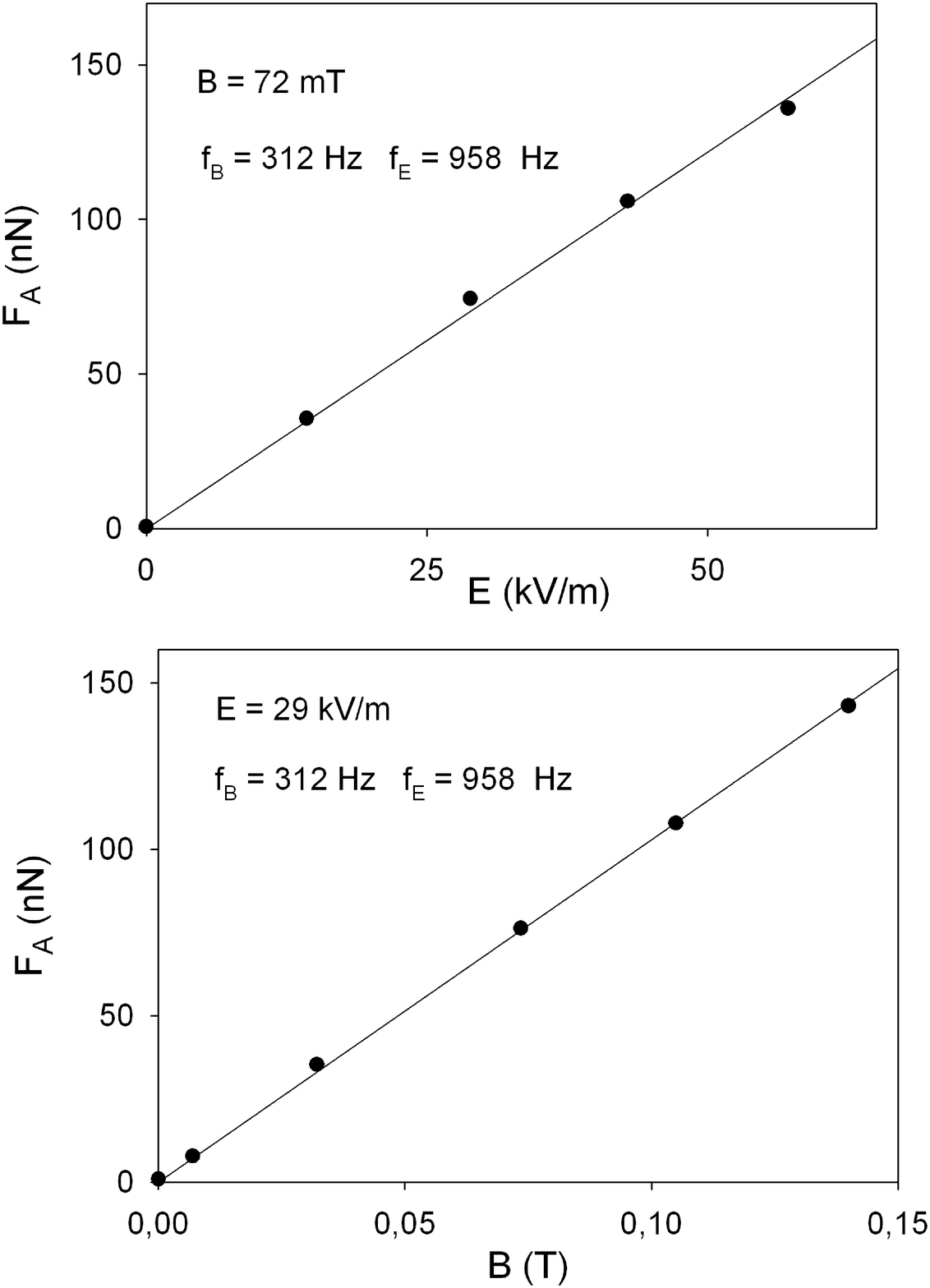}
}
\end{center}
\caption{Abraham force exerted on a slab of Y5V dielectric by crossed
oscillating electric and magnetic fields ($f_{i}=\omega _{i}/2\pi $).
Straight lines are linear fits to the data.}
\end{figure}

In this Letter we report the first observation of the \emph{intrinsic}
Abraham force on a dielectric induced by a time-varying magnetic field. Our
observations reveal the symmetry between electric and magnetic variations,
i.e we confirm $\mathbf{f}\propto \partial _{t}(\mathbf{E}\times \mathbf{B)}$%
, excluding the Peierls' version of radiative momentum, and in particular
confirming the Abraham and Nelson versions. We cannot discriminate between
the latter two, as $\mu =1$ in our case. On the theoretical side, our
finding supports the dominant role of pseudo-momentum in the controversy on
radiative momentum, and not the one of kinetic or conjugated momentum as was
suggested by other work \cite{Barnett}.

\begin{figure}
\begin{center}
\resizebox{0.4\textwidth}{!}{%
\includegraphics{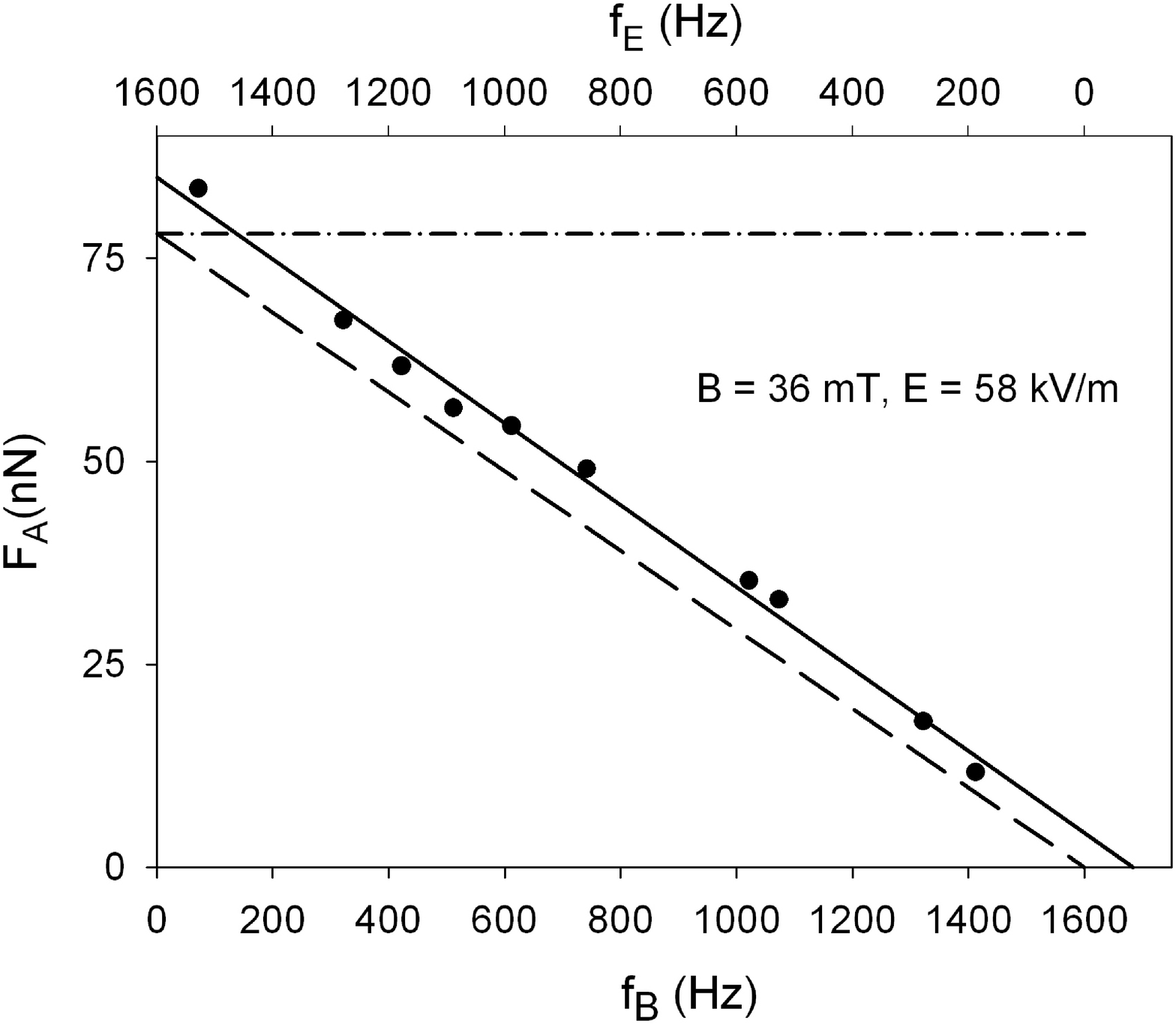}
}
\end{center}
\caption{Dependence of the Abraham force on a solidly contacted Y5V slab on
the frequencies of the electric and magnetic fields, whilst the sum
frequency and the amplitudes are kept constant. Solid line is a linear fit
to the data. Note that the intercept with the ordinate of the theoretical
lines (dashed, and dot-dashed, see text) is subject to a systematic
uncertainty of 5 \%.}
\end{figure}

A schematic view of our experiment is shown in Fig. 1. The sample consists
of a slab of Y5V dielectric recovered from a ceramic capacitor \cite{Murata}
(measured $\varepsilon _{r}=1,7\ 10^{5}$). It is covered on both faces by a
silverpaint electrode and mounted by means of an insulating spacer on a
piezoelectric bimorph cantilever \cite{Steminc}. The sensitivity of the
cantilever was determined by applying a known force to the sample and
measuring the charge generated by the cantilever with an electrometer. An
oscillating magnetic field $B\cos \omega _{B}t$ is applied perpendicular to
the electric field $E\cos \omega _{E}t$ inside the slab. The cantilever
signal is phase sensitively detected at $\omega _{B}+\omega _{E}$ and
corresponds therefore to a force on the sample proportional to $\mathbf{E(}t%
\mathbf{)}\times \mathbf{B(}t\mathbf{)}$. All other possible forces that
would result from field gradients\textit{\ }give a net zero contribution
because of the symmetry of our experimental geometry and appear at other
frequencies (see Eq. \ref{G}).

\begin{figure}
\begin{center}
\resizebox{0.4\textwidth}{!}{%
\includegraphics{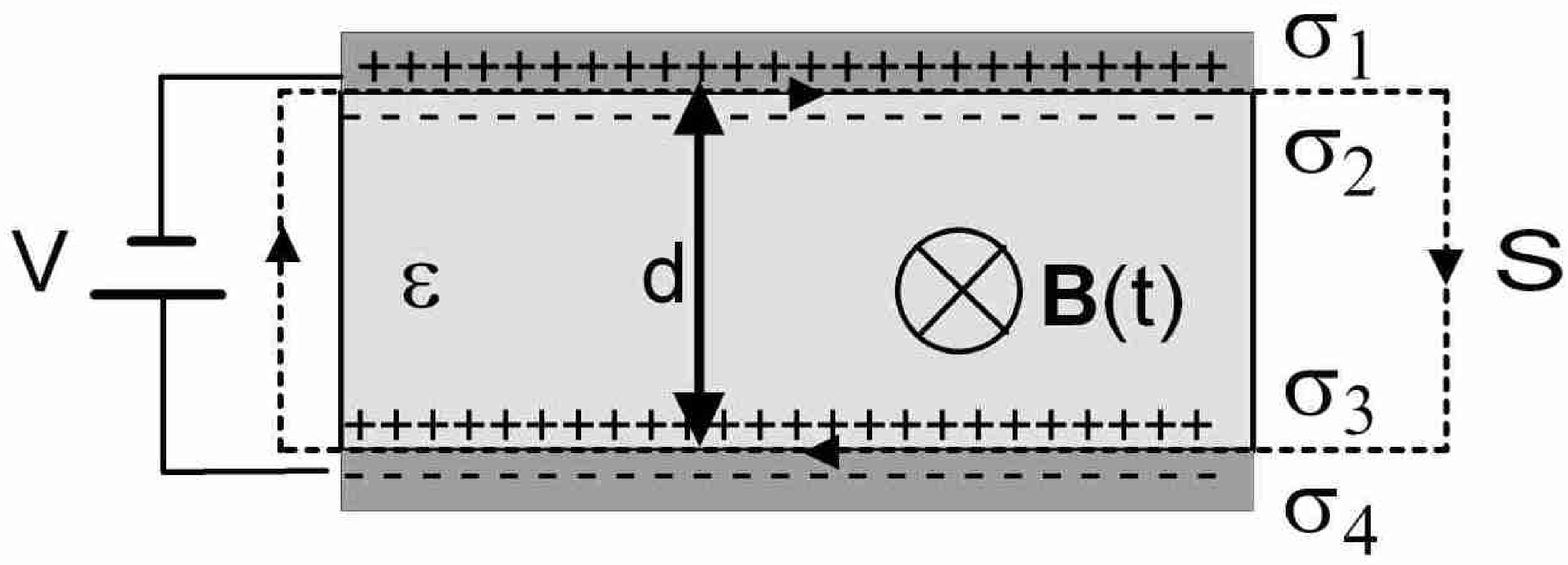}
}
\end{center}
\caption{Section of the sample + contacts, showing the different surface
charges.}
\end{figure}

A typical result is shown in Fig. 2. A clearly linear electric and magnetic
field field dependence of this force is observed, which we therefore
identify as the Abraham force. The linear dependence of \ $\mathbf{F}_{A%
\text{ }}=\mathbf{f}_{A}V$ , where $V$ is the sample volume, on the
frequency of the electric field, crossed with a static magnetic field, was
explicitly verified by Rikken and Van Tiggelen \cite{RikkenVanTiggelen},
i.e. $\mathbf{F}_{A\text{ }}\propto \mathbf{\partial }_{t}\mathbf{E\times B.}
$ To investigate the dependence of \ $\mathbf{F}_{A\text{ }}$ on the
magnetic field variation in time, we have varied $\omega _{B}$ whilst
keeping $\omega _{B}+\omega _{E}$ constant. This guarantees that the
sensitivity of our cantilever remains constant. The result is shown in Fig.
3, and shows clearly a strictly linear dependence on the electric field
frequency. So we find quantitative agreement with a prediction $\mathbf{f}%
_{A}=\varepsilon _{0}\left( \varepsilon _{r}-1\right) \mathbf{\partial }_{t}%
\mathbf{E\times B}$ (dashed line) and not at all with the predicted $\mathbf{%
f}_{A}=\varepsilon _{0}\left( \varepsilon _{r}-1\right) \partial _{t}\left(
\mathbf{E}\times \mathbf{B}\right) $ (dot-dashed line) ($\mu =1$ for Y5V).
From this, one could infer the absence of a contribution to $\mathbf{F}_{A%
\text{ }}$ of the form $\mathbf{E\times \partial }_{t}\mathbf{B}$. This
concurs with the findings by Walker et al \cite{Walker3}. The explanation
for the apparent absence of such a contribution is the almost complete
compensation of this contribution by an additional force on the electrodes.
Walker et al \cite{Walker5} have proposed a description for such a
compensation in terms of the Maxwell stress tensor. We propose a simpler
explanation, based on the different space charges present in the
sample-electrode system, as illustrated in Fig. 4.  It can be easily shown
for the surface charge densities $\sigma _{1}=\epsilon _{0}\varepsilon
V/d=-\sigma _{4}$ and $\sigma _{2}=-\epsilon _{0}(\varepsilon -1)V/d=-\sigma
_{3}$. The time varying magnetic field induces an electric field $\mathbf{E}%
_{ind}$\ that obeys%
\begin{equation}
\int_{S}\mathbf{E}_{ind}\cdot dl=\int \mathbf{\partial }_{t}\mathbf{B}\cdot
dS  \label{Induction law}
\end{equation}%
The Abraham force on the dielectric due to a time varying magnetic field
corresponds to the force exerted by $\mathbf{E}_{ind}$ on $\sigma _{2}$ and $%
\sigma _{3}$ but it will be almost completely compensated by the force
exerted by $\mathbf{E}_{ind}$ on $\sigma _{1}$ and $\sigma _{4}$ i.e. on the
electrodes. It follows that the force due to a time varying magnetic field
on the ensemble of electrodes plus dielectric is only $1/\left( \varepsilon
-1\right) $ of this force on the dielectric alone. It is therefore very
difficult to observe the $\mathbf{E\times \partial }_{t}\mathbf{B}$\
contibution to the Abraham force when the electric field is supplied by
electrodes that are fixed to the dielectric.

\begin{figure}
\begin{center}
\resizebox{0.35\textwidth}{!}{%
\includegraphics{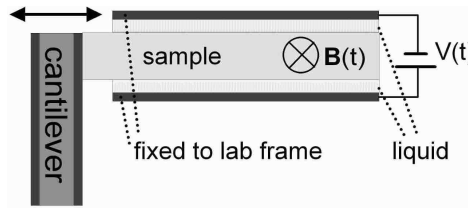}
}
\end{center}
\caption{Modified cantilever setup. Electrodes are fixed to the lab frame,
and electrical contact to the sample is provided by an ionic liquid.}
\end{figure}

In order to observe the intrinsic Abraham force on the dielectric, including
the $\mathbf{E\times \partial }_{t}\mathbf{B}$\ contribution, we have used a
configuration, where the electrodes are no longer rigidly connected to the
dielectric (Fig. 5). The electrodes are fixed in the laboratory frame, and
the sample is fixed to the cantilever, but otherwise free to move. The
electrical contact between sample and electrodes is provided by an ionic
liquid \cite{IL}. Now the ionic surface charges on the liquid side of the
liquid-sample interface (i.e. the equivalents of $\sigma _{1}$and $\sigma
_{4}$) that almost balance the surface charges on the sample, can move
freely in the liquid along the interface under the influence of $\mathbf{E}%
_{ind}$. They would therefore in \ the ideal case not exert any force on the
sample. In practice such ionic movement will partially be transferred to the
sample by the inevitable viscous drag of the liquid on the sample. Therefore
partial compensation of the Abraham force on the sample may still occur in
this configuration, but it should be much weaker than for the case of fixed
electrodes. Such a drag could also transfer a part of the Lorentz force $%
\mathbf{F}_{L\text{,}}$ experienced by the ionic current in the liquid, to
the sample. It can be easily shown that this Lorentz force is given by $%
F_{L}=IBl=2\pi f_{E}CVBl$, where $I$ is the current passing through the
liquid, $l$ is the total thickness of the two liquid layers, $C$ the
capacitance and $V$ the applied voltage.

The result of the measurement of the Abraham force with the liquid contacts
is shown in Fig. 6 as a function of the electric field frequency $f_{E}$,
where $f_{E}+f_{B}$ is kept constant. We clearly observe an Abraham force
that does not vanish for low $f_{E}$ as in Fig. 3. The full symbols are the
raw data, the open symbols are the data when corrected for the measured
resistive losses in the ionic liquid, which somewhat reduce the electric
field on the Y5V slab. Currently we do not have a conclusive explanation for
the small remaining negative slope. Combined with the earlier result that $%
F_{A}\propto f_{E}$ for a constant magnetic field \cite{RikkenVanTiggelen},
Fig. 6 proves that $F_{A}\propto $ $f_{E}+f_{B}\propto \partial _{t}(E\times
B\mathbf{)}$. For our parameters, the Lorentz force on the liquid is $F_{L}=$%
\textit{\ }$1,5$\textit{\ }$nN$\textit{\ }at\textit{\ }$f_{E}=100\ Hz$%
\textit{\ }and it has the same sign as $F_{A}$.\textit{\ }Its contribution
to the observed force on the sample can therefore be neglected at low $f_{E}.
$\ In the liquid contact configuration, it is difficult to accurately
determine the size of the contacted area of the sample and thereby of the
absolute value of the driving Abraham force density. From the measured
capacitance value we deduce a contacted area that would result in an Abraham
force of $15\ nN$, \textit{i}n reasonable agreement with the observed value.

\begin{figure}
\begin{center}
\resizebox{0.35\textwidth}{!}{%
\includegraphics{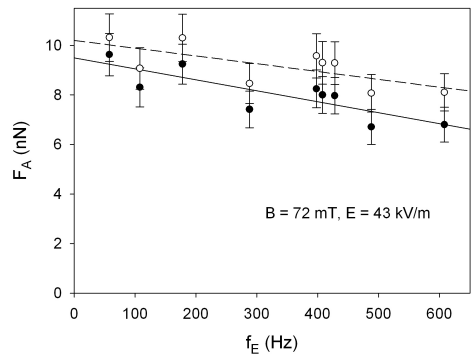}
}
\end{center}
\caption{Abraham force measured on a Y5V slab, contacted by an ionic liquid
as illustrated in Fig. 5. $f_{E}+f_{B}=690\ Hz$. Solid line is a linear fit
to the raw data (closed symbols). Dashed line is a linear fit to the
corrected data (see text).}
\end{figure}

In conclusion, we have confirmed the suggestion by Walker et al. \cite%
{Walker3,Walker4} that in solidly contacted dielectrics, the only measurable
contribution to the Abraham force is of the form $\mathbf{\partial }_{t}%
\mathbf{E\times B}$.\textit{\ }We have\textit{\ }provided a simple
explanation for this apparent absence of a $\mathbf{E\times \partial }_{t}%
\mathbf{B}$ contribution in terms of the compensating surface charge in the
electrodes. For a dielectric that is contacted by means of a conducting
liquid, we have observed the intrinsic Abraham force, which we find to be
proportional to $\partial _{t}\left( \mathbf{E}\times \mathbf{B}\right) $,
thereby explicitly verifying the Abraham and Nelson predictions of the
mechanical force density of electromagnetic fields in dielectrics. This
result greatly limits the arbitrariness in Eq. \ref{G}, leaving only the
possibility to assign terms derived from Maxwell's equations to either the
momentum or the stress tensor, but not to the force. In particular, our
results invalidate the Minkowski and Peierls versions of electromagnetic
momentum. In the solution proposed by Ref \cite{Barnett} the Abraham version
was associated with kinetic momentum, and Minkowski version with canonical
momentum. We have argued that the Nelson version is intimately related to a
third pseudo momentum. To finally discriminate between the two remaining
candidates, the Abraham and Nelson versions, one would have to discriminate
between $\left( \varepsilon _{r}-1/\mu _{r}\right) $ and $\left( \varepsilon
_{r}-1\right) $ as pre-factor, which will be the subject of future work.

We gratefully acknowledge Julien Billette for coil construction. This work
was supported by the ANR contract PHOTONIMPULS ANR-09-BLAN-0088-01.

\end{document}